# VISITOR SCHEDULE MANAGEMENT SYSTEM- AN INTELLIGENT DECISION SUPPORT SYSTEM

Srinivas Nidhra[1], Likith Poovanna[2] and Vinay Sudha Ethiraj[3]

School of Computing, Blekinge Tekniska Högskola, Karlskrona, Sweden
`nidhra.srinivas@gmail.com`

*ABSTRACT*

*Travelling salesman problem is a problem which is of high interest for researchers, industry professionals, and academicians. Visitor or salesman used to face lot of problems with respect to scheduling based on meeting top ranked clients. Even excel sheet made the work tedious. So these flaws propelled us to design an intelligent decision support system. This paper reports the problem definition we tried to address and possible solution to this problem. We even explained the project design and implementation of our visitor schedule management system.. Our system made a major contribution in terms of valuable resources such as time and satisfying high ranked clients efficiently. We used optimization via mathematical programming to solve these issues.*

## 1. INTRODUCTION

Travelling salesman problem is defined as finding the way for a salesman who must visit all the cities only once and then return back to where the salesman has started the journey. There are many algorithms such as genetic algorithm, heuristic algorithm to solve the travelling salesman problem. Many parameters can be considered to optimize the solution. Main parameters include distance, time and cost.

Our main aim is to intelligently design a support system for a visitor whose major concern is on management of time and meeting the customer priority wise.

## 2. PROJECT ANALYSIS

### 2.1. Background

There are 200 plus clients where each client has been assigned a rank ranging from 1 to 5 for being visited once or twice a year based on the rank. Our clients might own more than one terminal and are located in different part of city, country or continents. Some of them might be in the same country or even in same city.

The visitor needs to make decision on how and when to visit each client. Visitor has only 180 working days visitor needs to visit high ranked clients at least in the first 90 days. It is assumed that a trip from one city to another city or country to country takes maximum of one day and the meeting a client only takes half a day.
Direct communication is the usual way of doing business between the visitor and the clients. Direct communication means that every visitor either contacts the terminal operators or the visitor can directly contact the clients who own the terminal to set the appointments. Based on the responses from terminal operators the visitor will decide when he is going to visit each client and in which sequence. The visitor needs to have a confirmation of meeting either via phone, e-mail or fax from client's end before visitor plans the trip.
The aim of this project is to develop an intelligent decision support system that will help the visitor to use his valuable resources i.e. time and satisfying high ranked clients efficiently.

## 2.2. Problem Definition

Visitor face lot of problems while planning his visit based on the information present in excel sheet. Reasons being:

- Hard to interpret
- Incomplete information
- Irrelevant data
- Highly time consuming for maintaining and updating the excel sheet

Visitor cannot be everywhere at one point of time. Furthermore Visitor faces a huge problem of whom to visit at what time.

- Who to visit:
  Visitor cannot judge properly as to whom to visit. If the visitor starts off by meeting a high priority client. He then messes up his schedule either by visiting low priority client or he might end up wasting the day by not meeting any clients.

- When to visit:
  since visitor wants to visit all the high priority clients, they may not be available according to his schedule. Visitor then faces the problem of managing the time for rescheduling with the same client. Visitor might even end up in a dilemma as to which client to visit when many clients confirm at one shot.

## 2.3. Possible Solutions

The Intelligent decision support system we designed is an Expert System. We optimized the visitor's time management without compromising on meeting high priority clients on time as per confirmation. We have used genetic algorithm for visitor's scheduling management. Since our system is a rule based system, we had to follow certain rules to accomplish the task of managing the visitor's schedule.
The rules include:

- Half a day is the time given for visitor to visit 1 client
- A journey should take 1 day
- Working days of the visitor = 180 days
- Visitor has to visit the client if and only if the client confirms
- High rated clients has to be given first preference
- Clients placed in Rank 1 should be visited twice in a year
- Clients placed in Rank 2 and 3 should be visited once in a year
- If Twenty-foot equivalent unit > 500,000 , Place the client in Rank 1
- Clients with high ratings will be placed above the next higher rating clients during reconfirmation

### 2.3.1. System overview and algorithm details

#### 2.3.1.1 System Architecture:

Client information

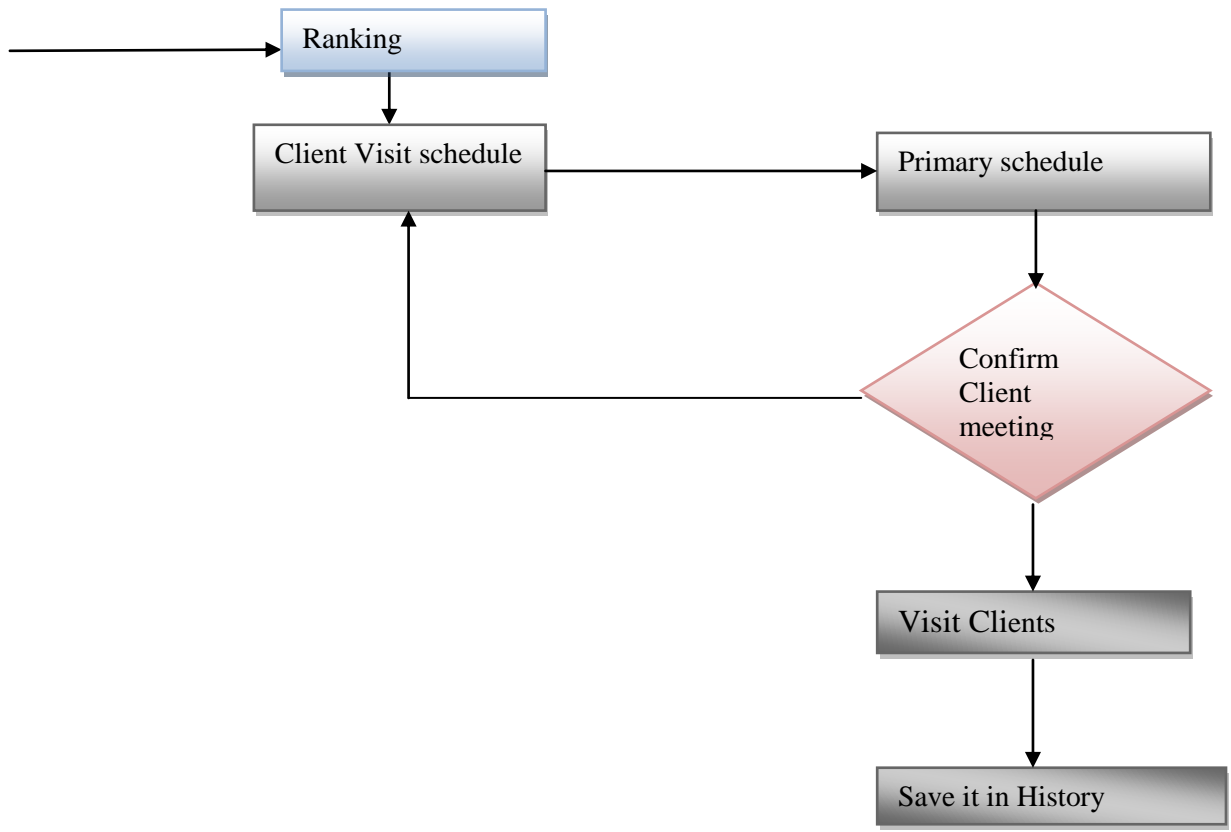

**2.3.1.2. Algorithm Details**

❖ Scheduling:

Since genetic algorithm is ideal for travelling salesman problem having more than 1000 cities. We used mathematical equation for developing a genetic algorithm on scheduling. Here visitor tries to get confirmation from clients present in the city having large number of top ranked clients. If clients deny his request for meeting, then scheduling algorithm intelligently regenerate the schedule with respect to the same city. If the number of confirmation turns out to be an odd number, then scheduling algorithm intelligently suggest next high ranked client for optimal utilization of time. Top ranked clients who belong to the same city and who are not confirmed will be placed above the city having number of top rated clients who belong to some other city lower than the former. Scheduling algorithm even works on suggesting rank to the visitor. The criteria the algorithm took for suggesting the rank were TEU (Twenty Foot Equivalent unit) because it is directly proportional to expansion of port implies clients is making high profit and should be ranked high. We considered the second criteria as country because visitor is keen on visiting those countries which are gaining importance in terms of profit.

## 3. PROJECT DESIGN

### 3.1. System Design
**The three-tier architecture of the visitor schedule management system**
The VSM application is designed as a web-based and server-side application that is partitioned in terms of application logic into three-tiers, Fig.1 Each layer has a different responsibility in the overall deployment and within each layer, and there can be one or more components. The layer partitioning is as follows. Presentation Tier contains components dealing with user interfaces and user interaction. The presentation layer of the web-based deployment could use HTML pages, Java Server Pages, and/or Java Applets. Middle Tier is composed of the Web Tier – JSP and Business Tier. The Web-Tier is the web server part that includes the web container and other protocols by the J2EE specifications. In the web container, servlets, JSP pages, filters, and web event listeners execute and may respond to HTTP requests from web clients. They may also be used to generate XML or other format data that is consumed by other application components.

The model-view-controller (MVC) paradigm was developed to map the input, processing, and output tasks to the graphical user interaction model. The *model* contains application data and behaviours, also providing an interface for the *view* and the *controller*. For each user interface, a *view* object is defined, containing the information about presentation formats, and is kept synchronized with the *model*'s state. Finally, the *controller* processes the user input and translates it into requests for specific application functionality. This separation reflects well the fact that Web applications may have different views, in the sense that it can be accessed through different clients, such as browsers, WAP clients, Web service clients, etc., with application data separated from its presentation. The existence of a separate module to handle user interaction, the *controller*, (or, more generally, interaction with other systems or user) provides better decoupling between the application behaviour and the way in which this behaviour is triggered. The advantage to partition the application into these logical layers is to isolate each tier. Thus, it should be possible to change the Presentation Tier while minimizing impacts on the business logic in the Application Tier or Data Tier.

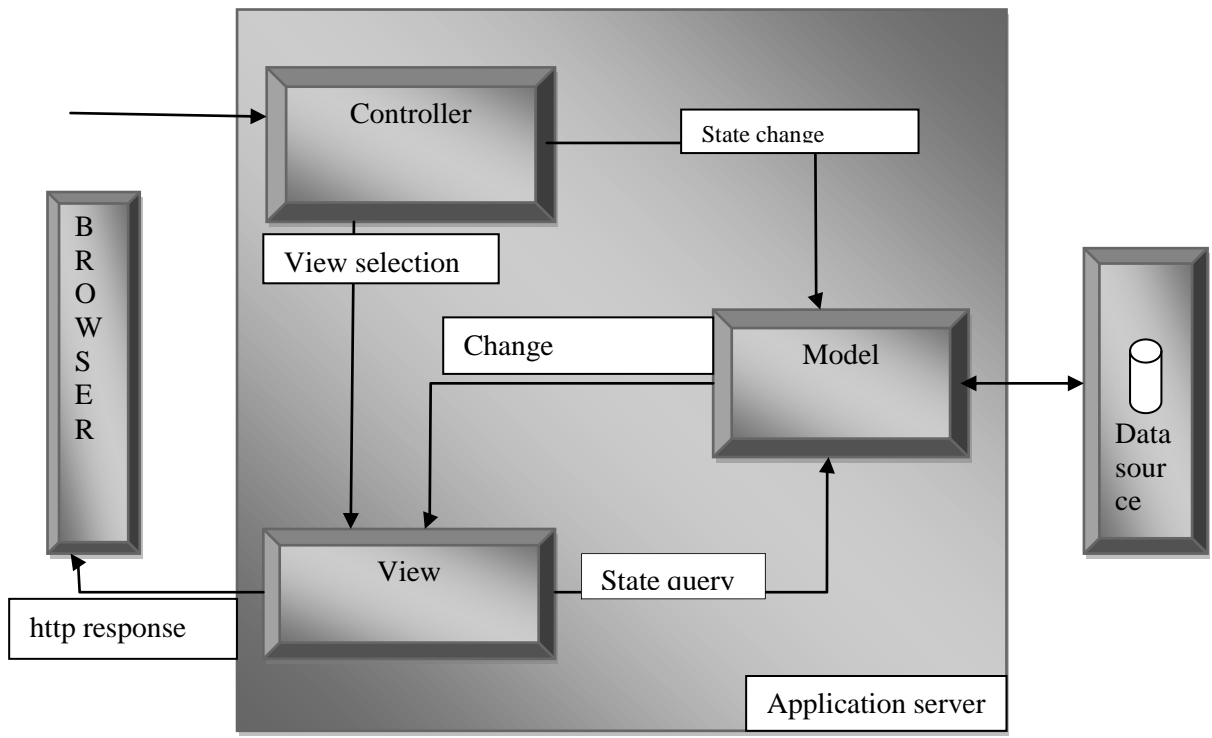

Fig. 1. MVC Architecture

### 3.2 Use Case Diagrams

Visual modelling is a communication tool that uses standard graphical notations for specifying, visualizing, constructing and documenting the software and/or business models, independently of the implementation language, [2]. UML is such a platform independent graphical standard used to define the visitors scheduling management architecture, the strategic reuse, system capabilities and application integration. The model of the system comprises several views that characterize it; use case and class diagrams (which define the functionality and the logical view of the system), state chart and activity diagrams (which define the behaviour of the system), sequence and collaboration diagrams (which define the interaction within the system), component and deployment diagrams (which define the implementation issues). It also has the following main capabilities: identify and design business objects and then map them to software components, partition services across a multi-tier model and/or architecture, design a distributed object architecture, code generation directly from the model, [3]. The conceptual design of the application is done using this visual modelling tool. The use case diagram for the VSM application is represented in Fig.2

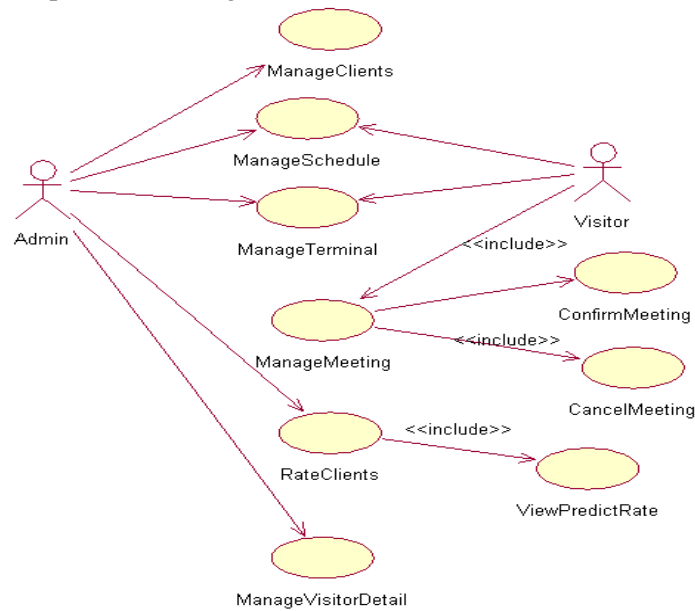

Fig. 2. Use case diagram for VSM application

UCDs have only 4 major elements: The actors that the system we are describing interacts with, the system itself, the use cases, or services, that the system knows how to perform, and the lines that represent relationships between these elements

| Usecase I | |
|---|---|
| ***Name*:-** | **Manage clients** |
| ***Description*:-** | This functionality helps to register, edit and remove client information. |
| ***Precondition*: -** | Client information must be available. |
| ***Post condition*:** | Client information is either registered or modified. |
| ***Basic Course of Action*:** | 1. User selects manage client<br>2. System provides the user with user management related functionalities(register, modify and remove)<br>3. If user selects register option<br>    a. System displays user registration form<br>    b. User fills client information and press submit button |

|  | c. The systems validate the data and register the user information.<br>4. If the user selects Modify Client information<br>    a. System displays edit form<br>    b. User modifies client information and press submit button<br>    c. The system validates the data and updates the user information.<br>5. If the user selects remove client option<br>    a. The system displays pop up window containing confirmation message.<br>    b. If the user agree to remove client detail again<br>    c. The system removes the client information from the system.<br>    d. Otherwise the system returns to the main page |
|---|---|
| *Alternate Course A*: |  |

| *Name***:-** | **Manage terminal** |
|---|---|
| *Description*:- | This functionality helps to register, edit and remove terminal information. |
| *Precondition*: - | There should be a client in the system |
| *Post condition*: | terminal information in the system |
| *Basic Course of Action***:** | 1. User selects manage Terminal option<br>2. System provides the user with terminal management related functionalities(register, modify and remove)<br>3. If user selects register option<br>    a. System displays terminal registration form.<br>    b. User fills terminal information and press submit button<br>    c. The system validates input data and register the terminal information.<br>4. If the user selects Modify terminal information<br>    a. System displays edit terminal form.<br>    b. User modifies terminal information and press submit button<br>    c. The system validates the data and updates the terminal information.<br>5. If the user selects remove terminal option<br>    a. The system displays pop up window containing confirmation message.<br>    b. If the user agree to remove terminal detail again<br>    c. The system removes the client information from the system.<br>    d. Otherwise the system returns to the main page |
| *Alternate Course A*: | |
| *Name*:- | **Manage visitor detail** |
| *Description*:- | This system functionality allows the user to register, modify and remove visitor information to the system. |
| *Precondition*: - | |
| *Post condition*: | Registered visitor. |
| *Basic Course of Action*: | 1. User selects manage visitor Information<br>2. System provides the user with visitor management related functionalities(register, modify and remove)<br>3. If user selects register option<br>    a. System displays visitor registration form<br>    b. User fills visitor information and press submit button<br>    c. The system validates the data and registers the visitor information.<br>4. If the user selects Modify visitor information<br>    a. System displays edit form<br>    b. User modifies visitor information and press submit button<br>    c. The system validates the data and updates the visitor information.<br>5. If the user selects remove visitor option<br>    a. The system displays pop up window containing confirmation message.<br>    b. If the user agree to remove visitor detail again<br>    c. The system removes the visitor information from the system.<br>    d. Otherwise the system returns to the main page |
| *Alternate Course A*: | |
| *Name*:- | **Calculate Rate** |
| *Description*:- | This system functionality is used to calculate the rate of client based on different parameters. |
| *Precondition*: - | Detail of client and terminal information should be available. |
| *Post condition*: | Rated of client. |
| *Basic Course of Action***:** | Rate client system component initiates rating of clients<br>1. The system providing the detail information of clients and terminal detail.<br>2. The system validates the available client data exists in the database<br>3. Calculates the rank of clients based on the formula. |
| *Alternate Course A*: | |
| *Name***:-** | **Rate clients** |
| *Description*:- | This functionality Enables the user to Rate the client based on different parameters (TEU, ...) |
| *Precondition*: - | The client should be in the system and has valid TEU value. |
| *Post condition*: | The Client will be Rated |
| *Basic Course of Action***:** | 1. User selects Rate Client option.<br>2. The system displays Client rating interface<br>3. The user select which Client to be rated. The system displays set rate manually or calculate option.<br>4. If the user select rate manually the system displays input box.<br>5. The user enters value and select submit option.<br>6. If The User selects Calculate Rate option the system initiates calculates rate for selected client.<br>7. System will save on the database |
| *Alternate Course A*: | |
| *Name***:-** | **Manage Schedule** |
| *Description*:- | This functionality allows the user to view client visit schedule by date or by client for the 90 and 180 days. In addition it will allow the user to update schedule based on primary meeting confirmation status of clients. |
| *Precondition*: - | The client detail and terminal information with ranking should be available |
| *Post condition*: | Detail schedule information. |
| *Basic Course of | 1. User selects manage schedule |

| *Action***:** | 2. The system displays preliminary General visit schedule. With the option confirm /un confirm meeting option. |
| --- | --- |
| | 3. If the user selects confirm option |
| | a. The system updates confirmation status of the client. |
| | 4. If the user selects un confirm meeting option |
| | a. The system regenerates schedule for that client and for the other clients form that day to the last working day. |
| | b. System will display updated schedule. |
| *Alternate Course A*: | |

### 3.3. Sequence Diagrams

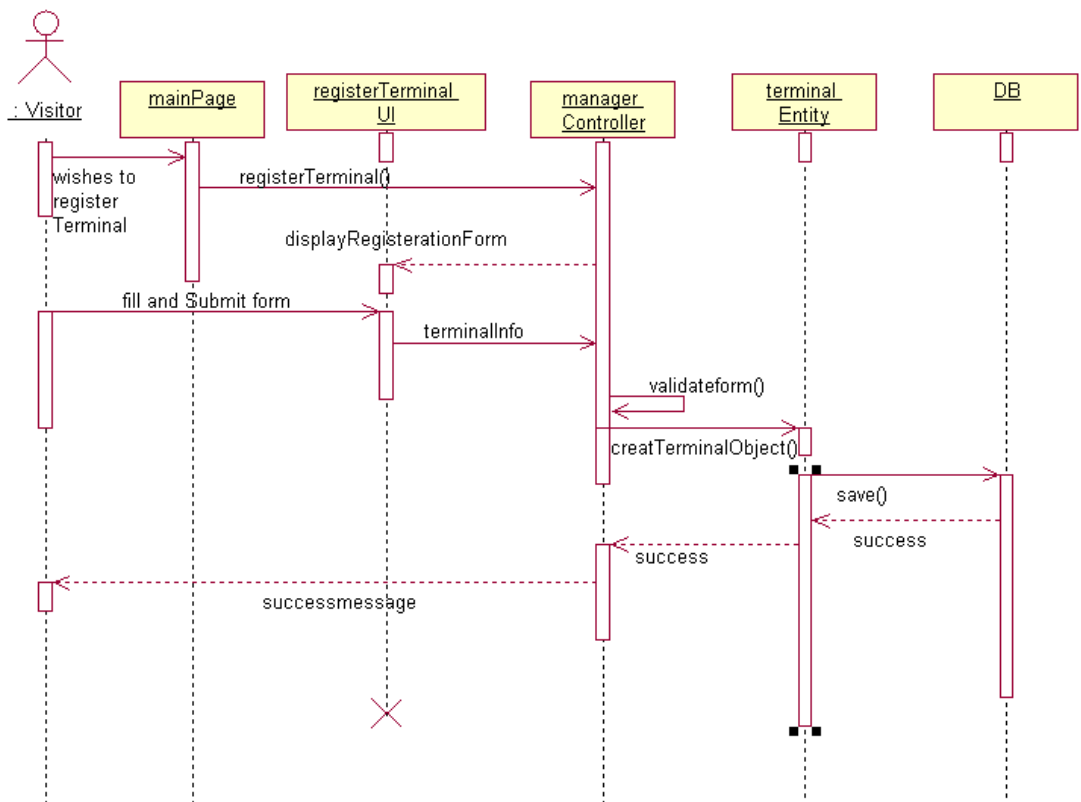

Fig. 3. Termitnal registration sequence diagram

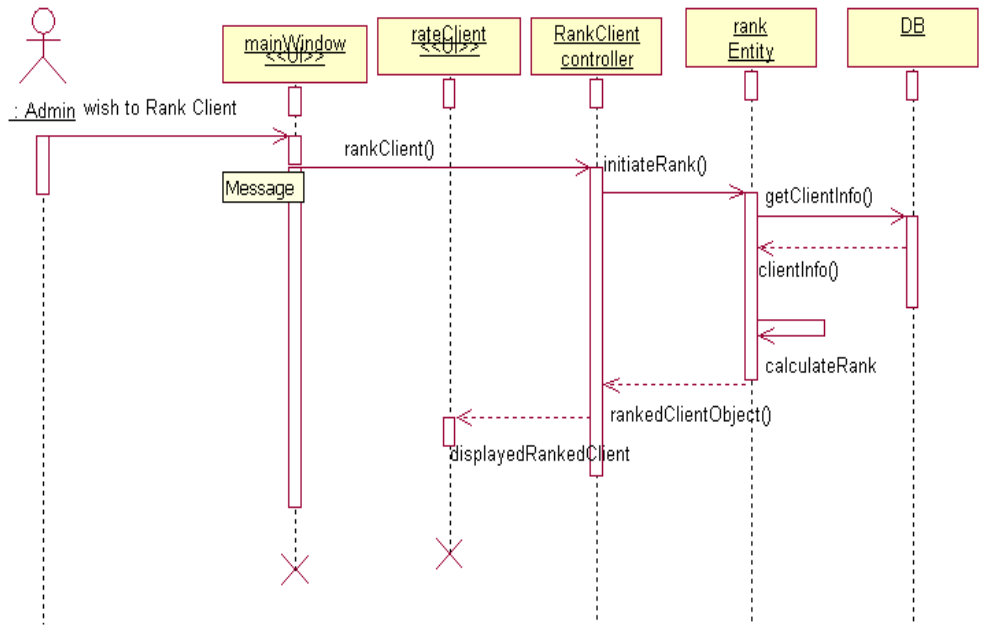

Fig. 4 Ranking clients sequence diagram

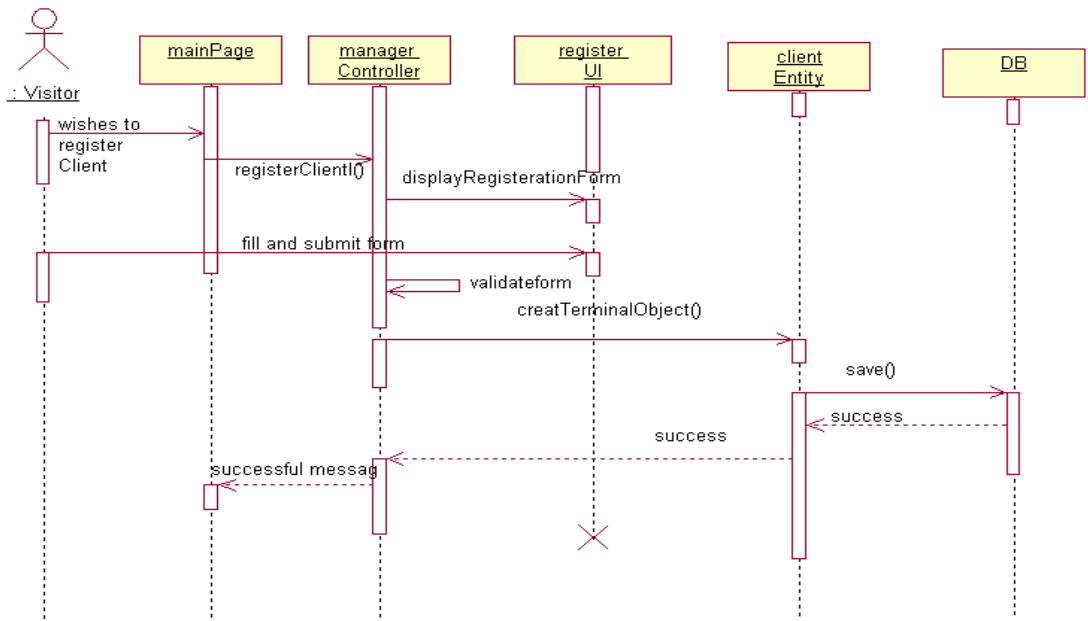

Fig. 5. Client registration squence diagram

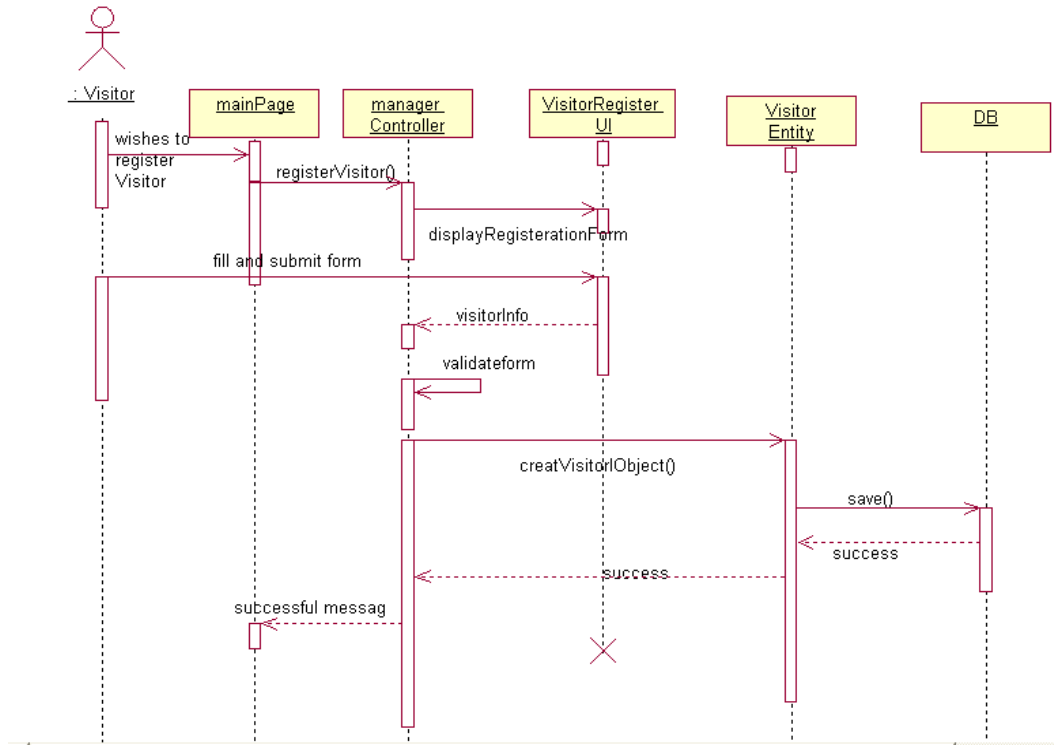

Fig. 6. Visitor registration sequence diagram

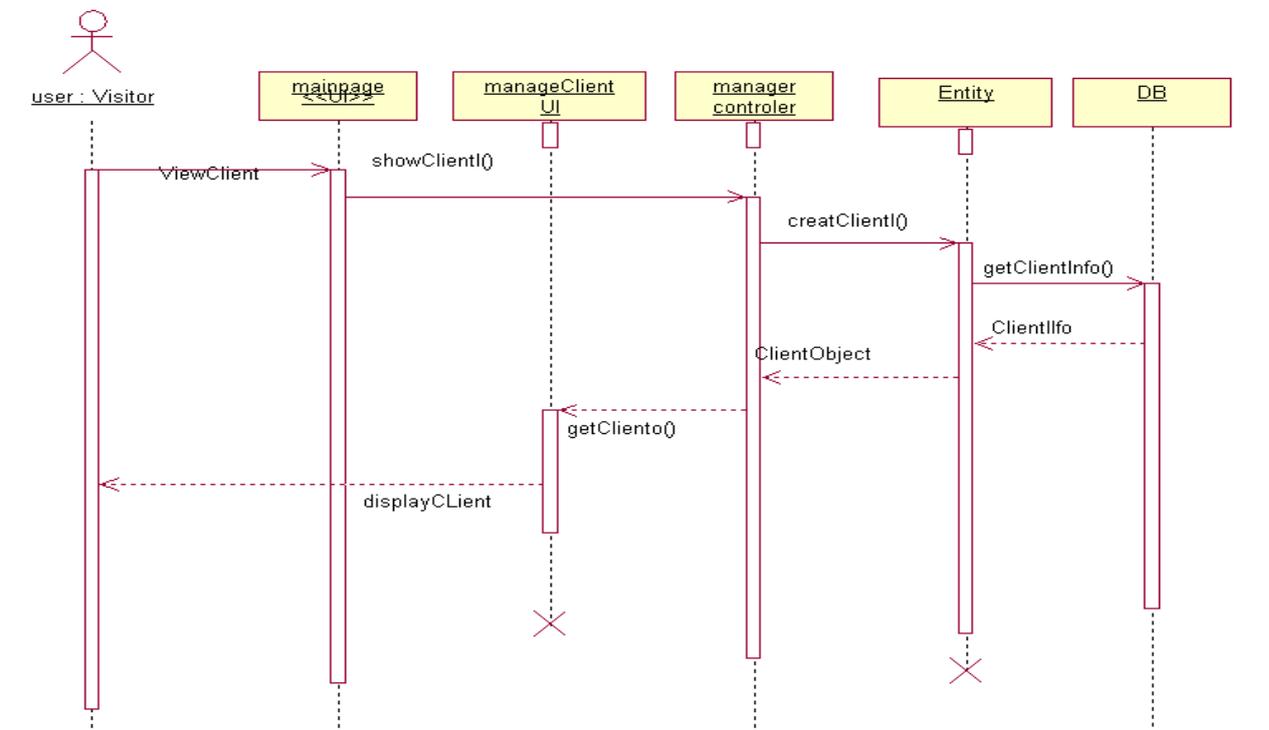

Fig. 7. View client sequence diagram

International Journal on Cybernetics & Informatics (IJCI) Vol.1, No.2, April 2012

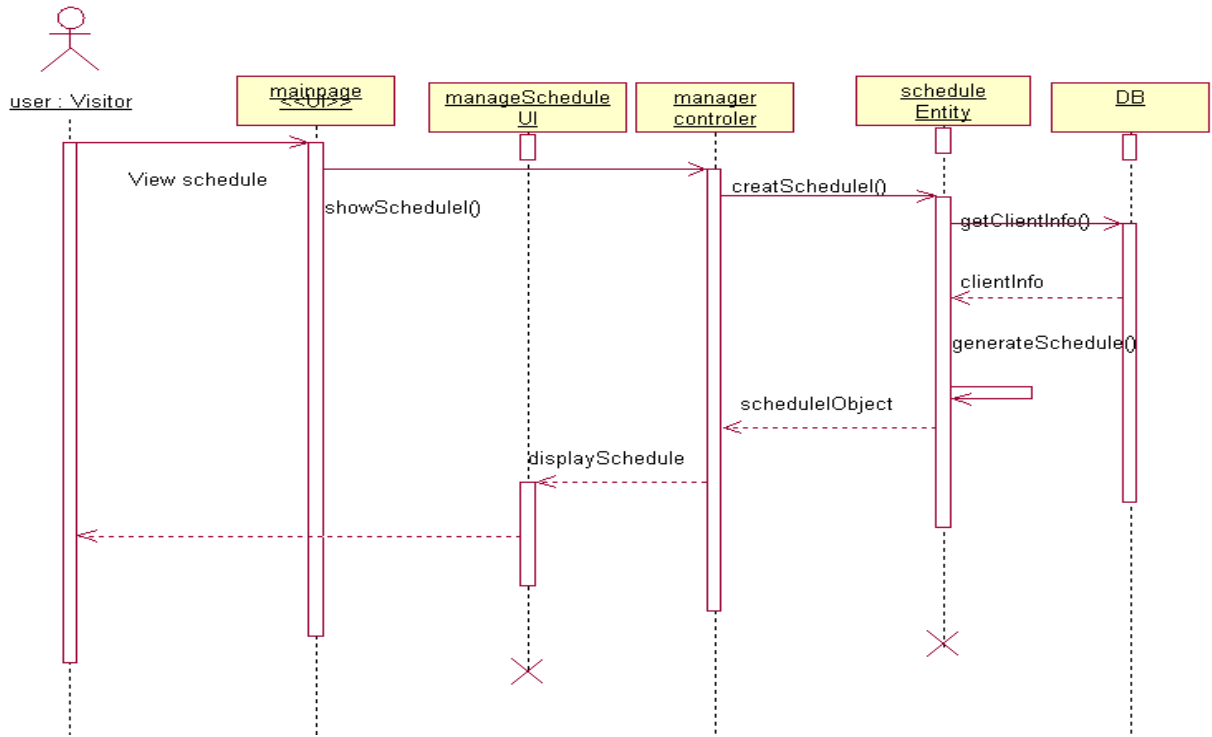

Fig. 8. View schedule sequence diagram

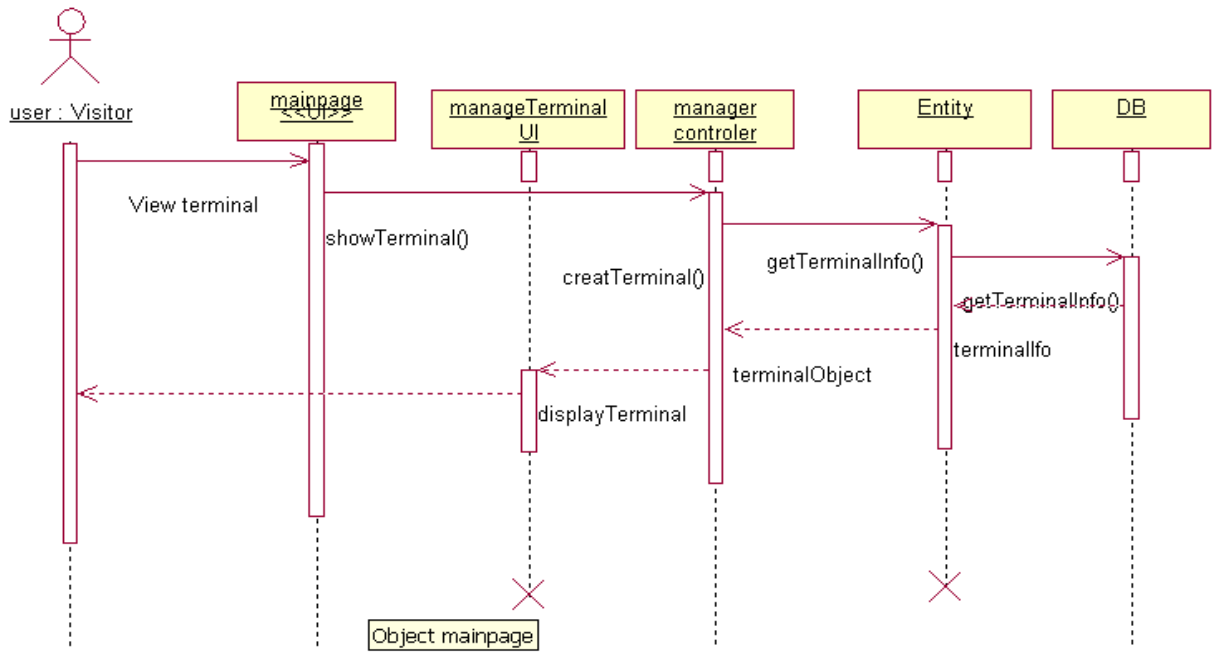

Fig. 9. View terminal sequence diagram

DOI: 10.5121/ijci.2012.1201

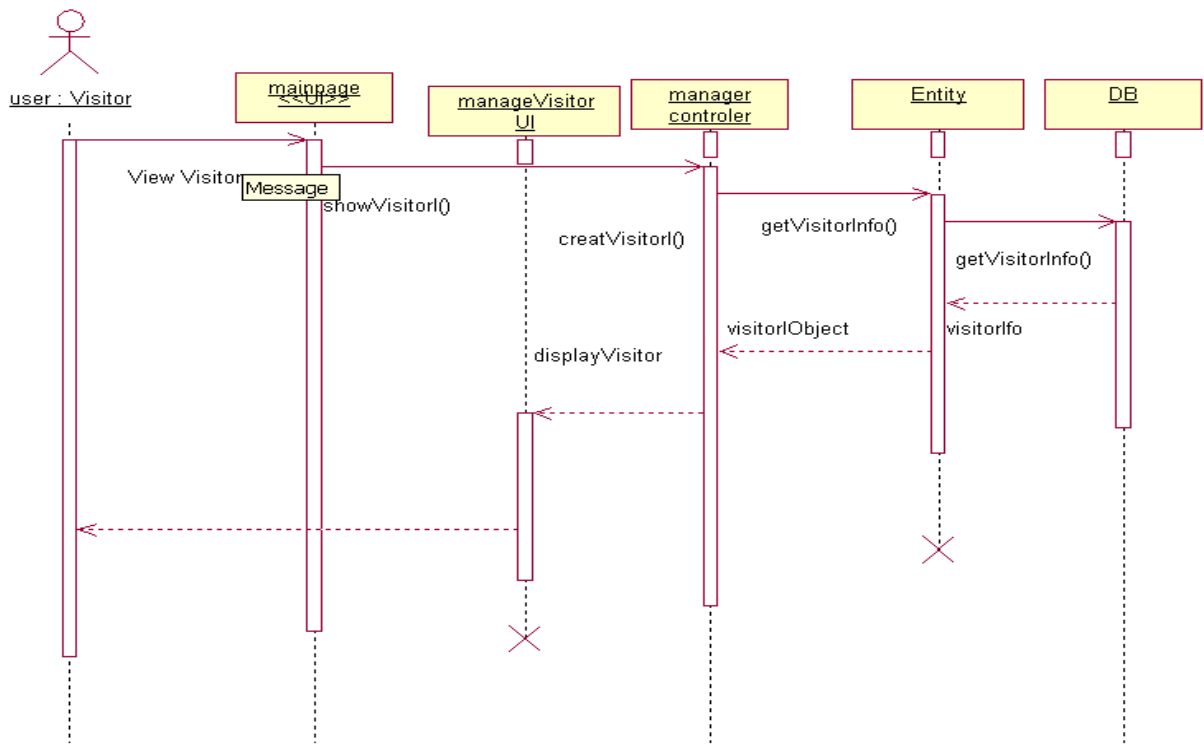

Fig. 10. View visitor sequence diagram

## 3.4 Class Relationship diagrams

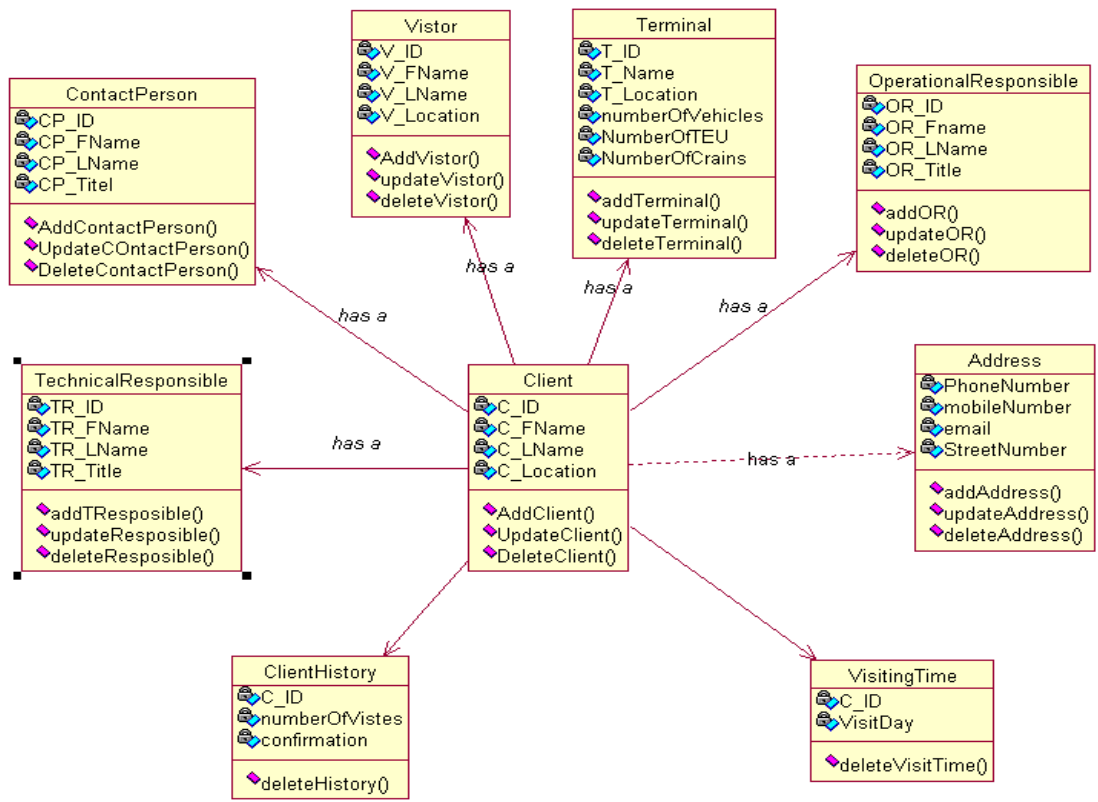

Fig. 11. Class diagram

## 4. PROJECT IMPLEMENTATION

### 4.1 Optimization via Mathematical Programming

For optimizing the visitor's meeting with preferably top ranked clients within the stipulated amount of time i.e. optimizing the visitor's schedule management requires an efficient mathematical equation. This implementation is depicted in fig 11.

To derive an equation, the parameters or variables which needs to be considered includes:

- ❖ c = city where client is located
- ❖ r = rank assigned to each client
- ❖ k = number of clients
- ❖ TTD = Total travel days
- ❖ TVD = Total visiting days
- ❖ 180 = Number of days allocated by visitor to meet clients in a year
- ❖ n = number of cities

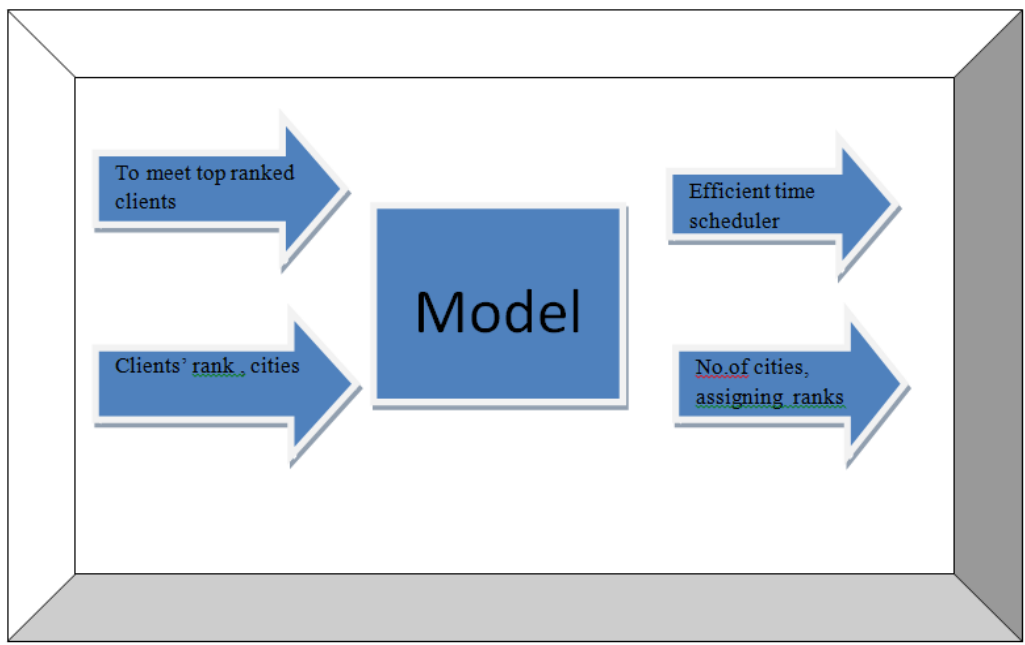

Fig. 12. Model Implementation

We know that, Visitor has 180 days to visit clients and travel from cities to cities. This can be depicted as
**TVD + TTD = 180**
We can visit only 2 clients per day (Rule) i.e.
**VD = $K_c$** where $K_c$ = 2
**TD = n-1** where n $\geq$ 2
We can easily analyze the total number of visiting days assigned in order to meet the clients' i.e.
**TVD = 180 – (n-1)**
But the problem in finding TVD using the above mentioned equation is that the equation does not consider giving priority to top ranked clients and also does not solve the problem of non-availability of these clients. Our main focus is on maximizing the visitor's visiting days in terms of meeting up top ranked clients based on confirmation; visitor might miss out certain cities because of 180 days constraint. So 'n' varies.

Number of days visitor is going to meet top ranked clients who belong to the same city is given by **$K_{rc}/2$**

Total visiting days (TVD) required by the visitor to meet all the clients who belong to the same city irrespective of ranks is given as

$$TVD = \sum_{r=1}^{5} K_r/2$$

Since our main aim is to make visitor meet clients based on rank, the above mentioned equation can be rewritten as

$$TVD = \sum_{r=1}^{5} \sum_{c=1}^{n} 1\, K_{rc}/2$$

For optimization purpose, certain conditions and protocols need to be followed namely,
- **$n \geq 2$** ( There has to be at least 2 cities for visitor to travel in order to meet the clients)
- **$K_{rc}$ % 2** if 1 then add 1 more client for effective utilization of time
- **$K_{rc} \geq 1$** There should be at least 1 client in each city
- **$K_{rc}\ \varepsilon$ {unvisited}** are those clients who are not visited yet

### 4.2 Process model

There are 4 stages in the process model namely:
1) **Case Retrieval**
2) **Case Reuse**
3) **Case Revision**
4) **Case Retention**

**Old/Previous case:** Schedule management carried out using excel sheet
**New case:** Schedule management carried out using Visitor Schedule Management system (Intelligent Decision Support System)

❖ Case Retrieval

It retrieves the best matching previous case. It has 3 tasks namely.
- Identification task: We carried out the set of problem descriptors which were relevant to our domain i.e. scheduling based on travelling salesman problem
- Matching task: We tried to match those set of cases that were very similar to the new case
- Selection task: Selection was carried out from those set of cases which best matches the new case we dealt with.

❖ Case Reuse

The case obtained from retrieval phase was used to figure out the differences among the past and the present case. We then transferred certain portion of the retrieved case for our future use.

❖ Case Revision

When the solution extracted from the reuse phase was not right, so revision of retrieved phase had to be carried out. Evaluation of case solution was performed since it was not successful we did not learn much from it. We then repaired the case solution using domain specific knowledge.

- ❖ Case Retention

Evaluation and repairing of case solution helped us to learn from success or failure.
- We retained the most desired part in our system. For instance some columns in excel sheet were transferred to database of our system
- We retained those portion in the form of data for our system
- We indexed the old case for later retrieval
- We finally integrated the new case into the memory structure

### 4.3 Intelligence

In this section we present how our system can be considered as an *intelligent* decision support system.

- ❖ **Generating schedule:**

Visitor tries to get confirmation from clients present in the city having large number of top ranked clients intelligently sorted using a mathematical equation. While asking for confirmation,
- If clients deny his request for meeting, then our system intelligently regenerate the schedule with respect to the same city.
- If the number of confirmation turns out to be a odd number, then our system intelligently suggest next high ranked client for optimal utilization of time.
- Top rated clients who belong to the same city and who are not confirmed will be placed above the city having number of top rated clients who belong to some other city lower than the former.
- ❖ **Suggesting Rank:**

Our system intelligently suggests rank based on:
- Variation (Increase/Decrease) in Twenty foot Equivalent Unit
- Client who is situated in a country which is gaining popularity from visitor's point of view.

## 5. COMMANKADS LIFECYCLE

The project management cycle comprises of 4 activities namely:
- ➢ Review
- ➢ Risk
- ➢ Plan
- ➢ Monitor

This process recurs in every cycle of the project.

**CommanKADS lifecycle 0**
- ❖ Step 1: Review

Review is the first step in project management lifecycle. In this phase, the status and main objective for upcoming cycles of the project is reviewed and establishe respectively.
- Objective

building an intelligent decision support system for the visitor who visits the client. Scheduling and effective time management is our main focus.

- Constraints:
  Client's availability
- Alternatives:
  Match the meeting of clients based on rankings and confirmation.
- Commitment:
  To increase the visitor's rapport with high ranked clients. This in turn leads to very good profit to the visitor.

❖ Step 2: Risk
- Identification of risks
- **Risk assessment is carried out**

| Risk | Affected Quality feature | Likelihood of occurrence | Severity of effect on project | Rank of risk | Counter measures |
|---|---|---|---|---|---|
| Non-availability of top ranked clients | Schedule functionality | High | High | 1 | Put in queue. Second ranked clients are approached intelligently |
| Satisfying visitor's need of meeting top ranked clients | Schedule functionality | Medium | High | 2 | Rescheduling algorithm to handle these needs |
| Missing good terminals | Ranking functionality | Low | Medium | 3 | Importance is given to countries of visitor's interest |

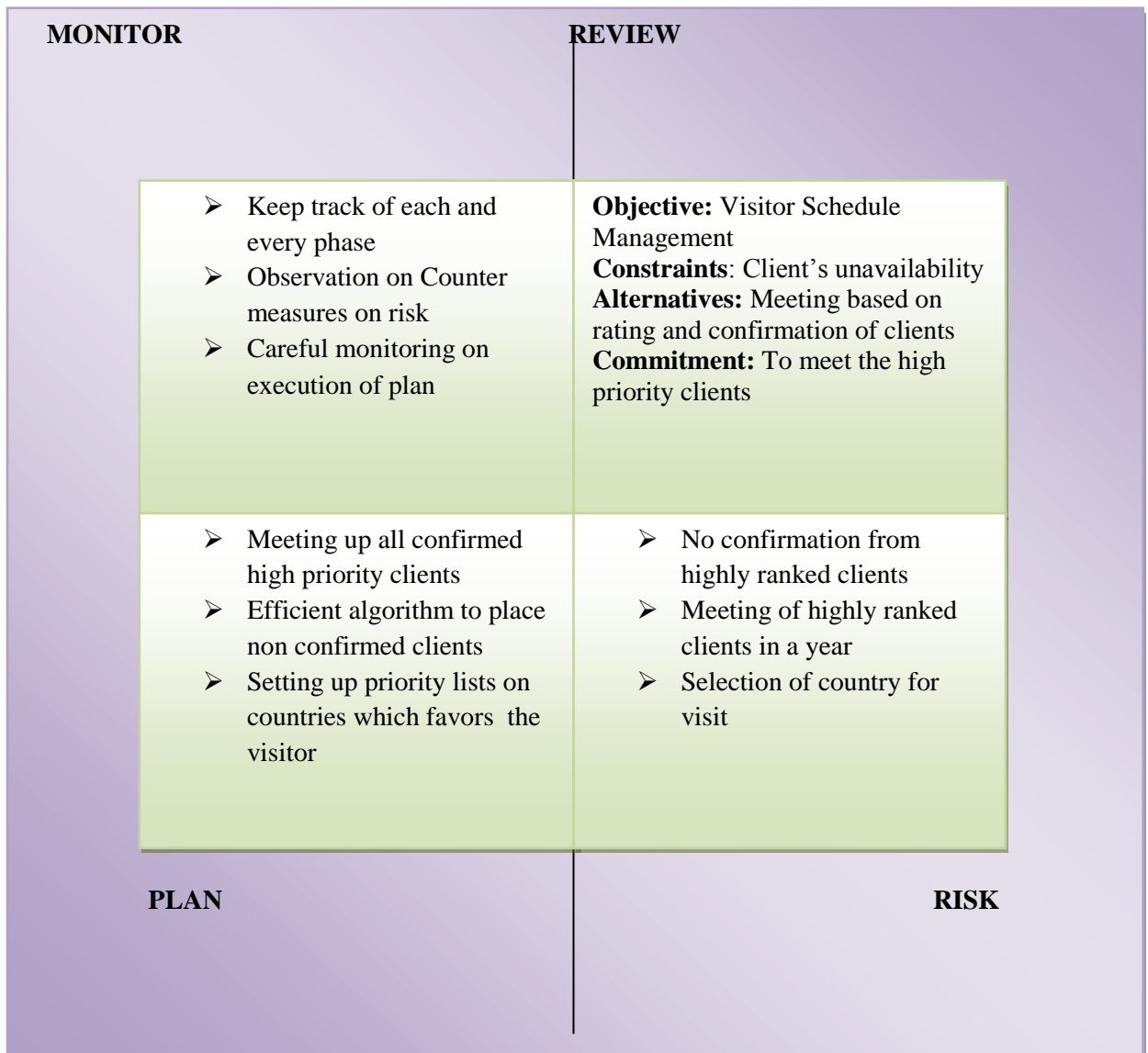

Fig.13. CommanKADS Lifecycle 0

- ❖ Step 3: Plan
- Satisfy visitor's urge of meeting top clients
- Effective rescheduling algorithm for non-availability of top clients
- To give importance to those countries which are of high interest with respect to visitor.
- ❖ Step 4: Monitor
- Keep track of each phase
- Identification of risks
- The plan was carefully monitored and implemented
- Plan for next cycle

**CommanKADS Lifecycle 1**

- ❖ Step 1: Review

International Journal on Cybernetics & Informatics (IJCI) Vol.1, No.2, April 2012Review is the first step in project management lifecycle. In this phase, the status and main objective for upcoming cycles of the project is reviewed and established respectively.

- Objective

Suggesting rank to the visitor for scheduling and effective time management based on updated ranking list

- Constraints:

Inclusion and exclusion of clients

- Alternatives:

Maintain a history chart for rank manipulation

- Commitment:

To increase visitor's efficiency of meeting clients with ease

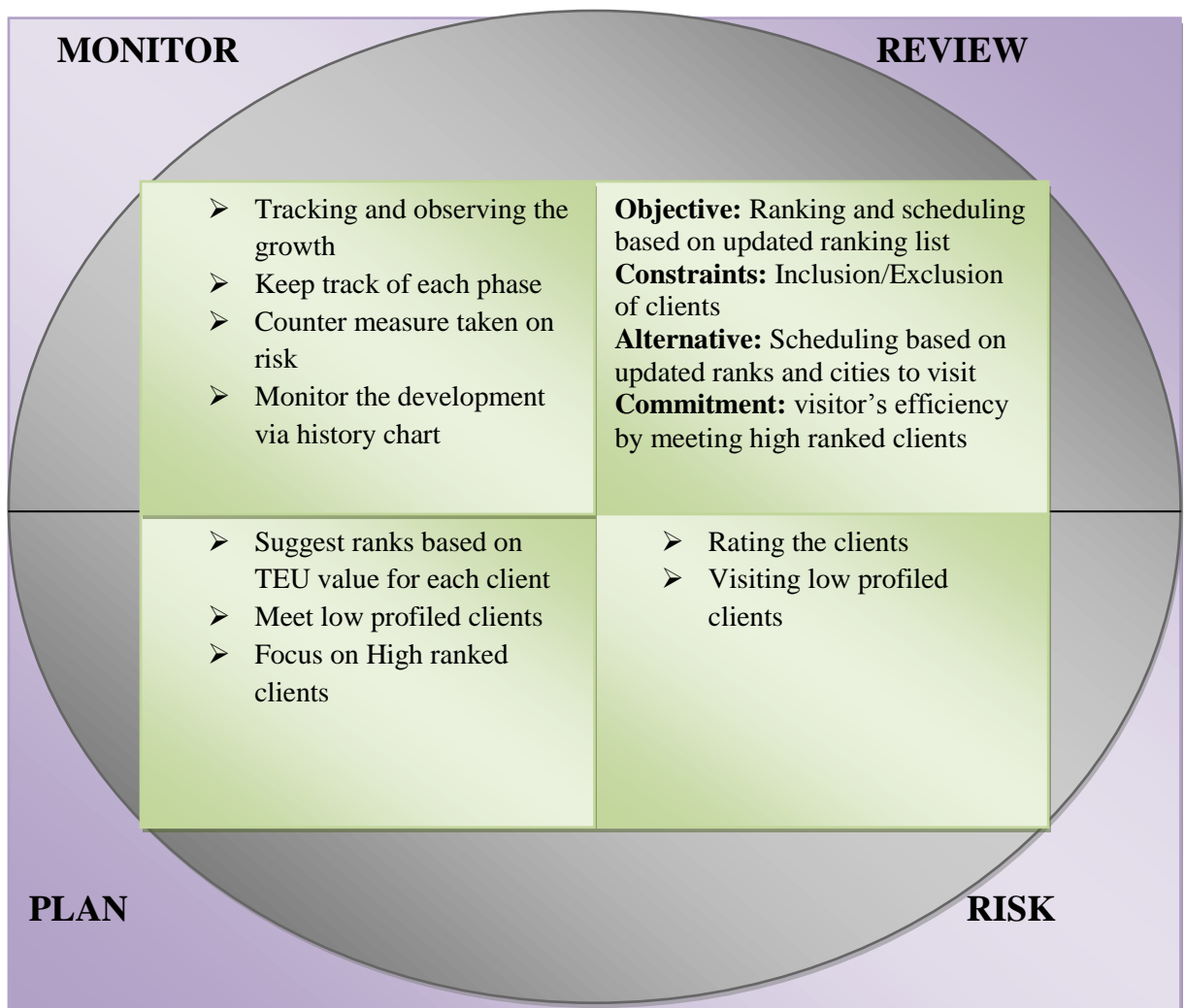

Fig.14. CommanKADS Lifecycle 1

- ❖ Step 2: Risk
- Identification of risks
- Risk assessment is carried out

19DOI: 10.5121/ijci.2012.1201

| Risk | Affected Quality feature | Likelihood of occurrence | Severity of effect on project | Rank of risk | Counter measures |
|---|---|---|---|---|---|
| Rating the clients | Ranking functionality | Medium | High | 1 | Algorithm to manipulate rank |
| Visiting low profiled clients | Ranking functionality | High | Low | 2 | Rule based algorithm on satisfying these clients |

- ❖ Step 3: Plan
- Manipulation and assigning new ranks to the clients based on criteria such as Twenty foot Equivalent Unit and countries of high interest for a visitor
- Meeting low profiled clients
- Effective schedule management based on ranking and country of visit.
- ❖ Step 4: Monitor
- Keep track of each phase
- Monitor counter measures taken on risks
- Tracking and observing the growth in visitor's daily work schedule
- Monitor the development visitor's time management via history chart

## 6. FUTURE DIRECTIONS

As a future work, we will be suggesting/planning the incorporation of each and every client's schedule by letting customer access our system in order to add their schedule. We would like to design it in such a way that our system has the ability to analyse different clients' schedule and reserve the meeting accordingly. When conflicts occur, the system will intelligently finds the available dates and sends a message to the client stating the available dates which in fact eases out the work from both the end.

## 7. CONCLUSIONS

Visitor cannot be everywhere at one point of time. Furthermore visitor faces a huge problem of whom to visit at what time. Visitor used excel sheet for scheduling his time for meeting the clients. Since excel sheet is hard to analyse visitor couldn't judge properly as to whom to visit. He might either meet low profiled client or end up wasting the day. Even visitor faced lot of problem of scheduling as to when to meet the high priority clients. So we designed a intelligent support system using optimization via mathematical modelling. Using this model we came up with the genetic algorithm which optimizes the visitor's schedule management keeping time as an integral element without compromising on meeting high priority clients as per confirmation.